\documentclass[a4paper,11pt]{article}
\pdfoutput=1 % if your are submitting a pdflatex (i.e. if you have
\usepackage{jinstpub} % for details on the use of the package, please
\usepackage{tabularx} % pretty tables
\usepackage[separate-uncertainty=true]{siunitx} % pretty SI units

\usepackage{placeins}

\usepackage{natbib}
\newlength\bibitemsep
\usepackage[moderate,mathspacing=normal,bibliography=tight,sections=tight]{savetrees} % tighter typesetting to save space

\graphicspath{{./fig/}}

\title{\boldmath Improved algorithms for determination of particle directions with Timepix3}

\author[a,b,1]{P.~Mánek,\note{Corresponding author.}}
\author[a]{B.~Bergmann,}
\author[a,c]{P.~Burian,}
\author[a]{D.~Garvey,}
\author[a,c]{L.~Meduna,}
\author[a]{S.~Pospíšil,}
\author[a]{P.~Smolyanskiy,}
\author[a,d]{and E.~White}

\affiliation[a]{Institute of Experimental and Applied Physics, Czech Technical University in Prague,\\Husova 240/5, 110~00 Prague, Czech Republic}
\affiliation[b]{Department of Physics and Astronomy, University College London,\\Gower Street, London, WC1E~6BT, United Kingdom}
\affiliation[c]{Faculty of Electrical Engineering, University of West Bohemia,\\Univerzitní 2795/26, 301~00 Pilsen, Czech Republic}
\affiliation[d]{School of Mathematics and Physics, Queen's University Belfast,\\University Road, Belfast, BT7~1NN, United Kingdom}

\emailAdd{petr.manek@utef.cvut.cz}

\abstract{%
Timepix3 pixel detectors have demonstrated great potential for tracking applications. With $256\times 256$~pixels, \SI{55}{\micro\meter} pitch and improved resolution in time (\SI{1.56}{\nano\second}) and energy (\SI{2}{\keV} at \SI{60}{\keV}), they have become powerful instruments for characterization of unknown radiation fields. A crucial pre-processing step for such analysis is the determination of particle trajectories in 3D~space from individual tracks. This study presents a comprehensive comparison of regression methods that tackle this task under the assumption of track linearity. The proposed methods were first evaluated on a simulation and assessed by their accuracy and computational time. Selected methods were then validated with a real-world dataset, which was measured in a well-known radiation field. Finally, the presented methods were applied to experimental data from the Large Hadron Collider. The best-performing methods achieved a mean absolute error of~\ang{1.99} and~\ang{3.90} in incidence angle~$\theta$ and azimuth~$\varphi$, respectively. The fastest presented method required a mean computational time of~\SI{0.02}{\pico\second} per track. For all experimental applications, we present angular maps and stopping power spectra.
}

\keywords{%
Analysis and statistical methods;
Data processing methods;
Data reduction methods;
Pattern recognition, cluster finding, calibration and fitting methods
}

\arxivnumber{2111.00624}

\proceeding{22$^{\text{nd}}$ International Workshops on Radiation Imaging Detectors\\
  June 27 - July 1, 2021\\
  Ghent, Belgium}

\begin{document}
\maketitle
\flushbottom

%%%%%%%%%%%%%%%%%%%%%%%%%%%%%%%%%%%%%%%%%%%%%%%%%%%%%%%%%%%%%%%%%%%%%%%%%%%%%%%%%%%%%%%%%%%%%%%%%%%%%
%% BODY ORGANIZED BY SECTIONS
%%%%%%%%%%%%%%%%%%%%%%%%%%%%%%%%%%%%%%%%%%%%%%%%%%%%%%%%%%%%%%%%%%%%%%%%%%%%%%%%%%%%%%%%%%%%%%%%%%%%%

\section{Introduction}
\label{sec:intro}

Hybrid active pixel detectors of the Timepix family have shown considerable potential for applications in tracking and radiation field characterization. Their capability for particle identification has been widely studied since the Medipix2 chip that permitted rudimentary classification of tracks on the basis of morphological features~\cite{holy2008pattern}. Its successor Timepix enabled energy measurements in Time-over-Threshold counting mode, opening possibilities for including spectroscopic information in feature sets. Fitting linear trajectories to tracks allowed calculation of stopping power $\mathrm{d}E/\mathrm{d}X$, which proved to contain valuable information about particle species~\cite{Manek2018_CTU}. Furthermore, estimated linear trajectories allowed directional studies of complex radiation fields such as the ATLAS or the MoEDAL Experiment at CERN~\cite{Manek2019_CTDWIT_proc,bergmann2019characterization}, or in space~\cite{ruffenach2021new}. Accuracy of trajectory estimation clearly affects the utility of track features such as $\mathrm{d}E/\mathrm{d}X$, which in turn impacts the performance of any classifier operating on them. This motivates a search for accurate directional estimation methods for Timepix chips, and in particular for the latest addition to the family, the Timepix3~\cite{poikela2014timepix3}.

Similarly to Timepix, Timepix3 comprises $256\times 256$~pixels with \SI{55}{\micro\meter}~pitch and different acquisition modes. In addition to increased energy and time resolution over Timepix (\SI{2}{\keV} at \SI{60}{\keV} and \SI{1.56}{\ns}, respectively), Timepix3 chips offer several key advantages that are desirable for tracking applications. Firstly, they support measuring both time and energy simultaneously in each pixel. Secondly, they offer a data-driven readout scheme that permits data to be taken continuously rather than in frames with a constant exposure time. In this scheme, hit pixels are reported immediately while the rest of the chip remains sensitive as they are read out, reducing dead time. Finally, Timepix3 has demonstrated a unique 3D reconstruction capability that leverages nanosecond-precise time information to assign relative depth coordinates to hits~\cite{bergmann20173d,bergmann20193d}, presenting an opportunity for development of more sophisticated tracking techniques~\cite{bergmann2020particle,acharya2021timepix3}.

Focusing on one of the essential steps leading to feature extraction for particle identification in Timepix3, this work presents a variety of regression methods for determination of linear trajectories in 3D. First, section~\ref{sec:meth} defines the methods as well as metrics and datasets used for their evaluation. Second, section~\ref{sec:comp} presents evaluation results, compares presented methods and justifies the selection of methods that are used in the following sections. Third, section~\ref{sec:exp} describes application of the selected methods to experimental data, and discusses the obtained results. And last, section~\ref{sec:conclusion} summarizes the contribution and suggests possible directions for future development.

\section{Methodology}
\label{sec:meth}

In data-driven mode, used in the present study, Timepix3 transmits a continuous stream of hits. Using temporal and spatial coincidence requirements, hits are grouped into so-called tracks that correspond to single particle interactions in a silicon sensor. With individual tracks as inputs, we proceed to formulate the task of directional estimation as a conventional regression problem.

Given a track that consists of multiple hits (each containing X, Y coordinate, a timestamp and deposited energy), we aim to estimate two spherical angles that determine the 3D direction of the corresponding particle's trajectory: (1)~azimuth~$\varphi$ that describes orientation of the trajectory in the incidence plane of the sensor and (2)~incidence angle~$\theta$ that the trajectory closes with the normal of that plane.

%%%%%%%%%%%%%%%%%%%%%%%%%%%%%%%%%%%%%%%%%%%%%%%%%%%%%%%%%%%%%%%%%%%%%%%%%%%%%%%%%%%%%%%%%%%%%%%%%%%%%
\subsection{Regression methods}

To offer a broad comparison of techniques that may be used to tackle the task of directional estimation, we propose a variety of regression methods. These methods represent a balance between approaches that were previously used with success and ones that have been only recently enabled by the new features of Timepix3. In this work, we formalize them as algorithms that take a single track as input, and produce a pair of predicted regression parameters~$(\varphi,\theta)$.

The proposed methods can be split in two distinct categories according to their implementation. While some estimate both regression parameters directly from data, others operate in two stages. The first stage, which usually differs between methods, selects a pair of hits from the track. The second stage, common to all such methods, calculates both regression parameters analytically under the assumption that selected hits represent particle's entry and exit points in the sensor. What follows is a brief explanation of each method.

\begin{description}
\item[Left Lower-most, Right Upper-most (LLM)]
This method selects a pair of hits with the smallest and the largest X, Y~coordinate, respectively. The selected hits are assumed to be the entry and exit points of a linear trajectory.

\item[LLM-Improved]
Extending the LLM method, this algorithm considers dimensions of the track's bounding box. If its height exceeds its width, the left lower-most hit is determined by minimizing along the Y~axis and then along the X~axis, otherwise the search order is reversed. Analogous logic is applied to selection of the right upper-most hit.

\item[Time Min-Max]
This method selects a pair of hits with the smallest and the largest timestamp. The selected hits are assumed to be the entry and exit points of a linear trajectory.

\item[Time Gradient]
This method assigns new attributes to each hit: (1)~depth coordinate reconstructed from timestamps~\cite{bergmann20173d}, and (2)~distance from the hit with the smallest timestamp. With these attributes, least squares method is used to fit a 2D~line, from which regression parameters are calculated analytically.

\item[Line Fit]
This method estimates regression parameters independently. Azimuth~$\varphi$ is calculated from the slope of a 2D~line that is fitted through all hits using least-squares algorithm that weighs individual hits by their energy. Incidence angle~$\theta$ is determined from eigenvectors of squared matrix of 3D~hit coordinates, where depth coordinates were previously reconstructed from hit timestamps~\cite{bergmann20173d}.

\item[Time-weighted]
This method finds boundary points of a linear trajectory using weighted means of all hit coordinates in the track. Prior to calculation, the smallest timestamp $\min_{j=1}^n t_j$ is subtracted from all hit timestamps $\{t_i\}_{i=1}^n$ in the track. While the entry point is determined using weights~$t_i/\sum_{j=1}^n t_j$, the exit point calculation uses weights~$(\max_{j=1}^n t_j - t_i)/\sum_{j=1}^n t_j$. Locations of the entry and exit point are then used to calculate regression parameters analytically.

\item[Random Sample Consensus (RANSAC) \cite{RANSAC}]
Commonly used in computer vision, this method iteratively samples pairs of hits uniformly at random. For each pair, a linear model is analytically determined and a set of inliers (hits that are sufficiently close to fitted model) is found. Following a number of iterations expected to achieve a viable sampling with parametric probability $p$, the model with the largest inlier set is viewed as a linear trajectory, from which regression parameters are calculated.
Since the value of the $p$~parameter controls a possible trade-off between speed and accuracy, we evaluate two variants of this algorithm, one labeled RANSAC-Fast with $p=0.90$, and another labeled RANSAC-Prec with $p=0.9999$.

\item[Neural Network]
This method evaluates two artificial neural networks with a multi-layer perceptron architecture, one for each regression parameter. At the input, results of other methods (Line Fit, LLM-Improved) are propagated through 2~hidden layers containing 10~neurons with ReLU~activation\footnote{Rectified linear unit (ReLU) activation function is defined as~$f(x)=\max\{0,x\}$ \cite{ReLU}.}. The final output layer contains a single neuron with linear activation that generates predicted values.

\item[Hough Transform~\cite{HoughLineTransform}]
This method populates a 2D~phase space indexed by hypothetical model parameters~$(\theta,r)$ that were used in previous studies~\cite{Manek2018_CTU}. Hits of the track are enumerated such that each hit votes for all consistent model instances, increasing their weight relative to others by the amount proportional to deposited energy. Finally, local extrema are identified that correspond to the most likely parameter values, which are used to calculate regression parameters $(\varphi,\theta)$ analytically.
Since in implementation phase space is represented as a 2D~histogram, its bin widths~$(\Delta\theta,\Delta r)$ control trade-off between accuracy and tractability. Similarly to RANSAC, we therefore consider two variants of this algorithm that differ in bin widths: (1)~Hough-Fast with $(\ang{1.72},\SI{1.65}{\micro\meter})$, and (2)~Hough-Prec with $(\ang{0.57},\SI{55}{\micro\meter})$.
% 0.03rad slope, 0.03px offset
% 0.01rad slope, 1px offset

\item[Decision Tree~\cite{scikit-learn,DecisionTrees}]
This method evaluates two decision tree regressors, which are used similarly to the Neural Network. Specifically, each regression parameter is predicted independently by a previously trained model that takes inputs from predictions of other methods (Line Fit, LLM-Improved). While the decision tree that predicts azimuth~$\varphi$ permits maximum depth of~11 nodes and requires that at least 5~samples are present in every leaf, the tree that predicts incidence angle~$\theta$ permits trees of depth at most~8 and requires at least 2~samples in each leaf.

\end{description}

%%%%%%%%%%%%%%%%%%%%%%%%%%%%%%%%%%%%%%%%%%%%%%%%%%%%%%%%%%%%%%%%%%%%%%%%%%%%%%%%%%%%%%%%%%%%%%%%%%%%%
\subsection{Metrics}

In line with conventional methodology, the presented methods are primarily evaluated by their accuracy. To this end, we report the following figures for each parameter $x\in\{\varphi,\theta\}$: (1)~mean absolute error~$\Delta$, (2)~outlier percentage~$O$, (3)~coefficient of determination~$R^2$. These metrics are defined as%
\begin{align}
    \Delta = n^{-1}\sum_{i=1}^{n} \left|x_i - \hat{x}_i\right|,
    \qquad
    O = n^{-1}\sum_{i=1}^{n} \mathbb{I}\left[\left|x_i-\hat{x}_i\right|>3\sigma_x\right],
    \qquad
    R^2=1 - \frac{\sum_{i=1}^{n}\left(x_i-\hat{x}_i\right)^2}{ \sum_{i=1}^{n}\left(x_i-\overline{x}\right)^2}.
\end{align}
Here true values and predictions are denoted $\left\{x_i\right\}_{i=1}^{n}$ and $\left\{\hat{x}_i\right\}_{i=1}^{n}$, respectively. Furthermore, $\overline{x}$ labels arithmetic mean calculated from true values $x_i$, $\mathbb{I}$ is a binary indicator variable and $\sigma_x$ denotes the standard deviation of true values $x_i$.

Due to the high bandwidth of Timepix3 that can theoretically reach 80\,Mhits/s, it is desirable to seek regression methods that minimize execution time, so as to achieve tractability in real-time during data acquisition. To this end, we calculate mean time per track~$\overline{t}$ as~$t/n$, where $t$ is the total wall time required to process $n$ tracks. All time measurements presented in this work were performed on a PC\footnote{Dell Latitude 5501 Business Laptop~\cite{DellLatitude5501}.} with Intel\textsuperscript{\tiny\textregistered{}} Core\texttrademark{} i7-9850H CPU and 16~GB RAM.

%%%%%%%%%%%%%%%%%%%%%%%%%%%%%%%%%%%%%%%%%%%%%%%%%%%%%%%%%%%%%%%%%%%%%%%%%%%%%%%%%%%%%%%%%%%%%%%%%%%%%
\subsection{Datasets}

For the purposes of this work, evaluation was performed with both experimental and simulated datasets. All such datasets were generated by Timepix3 chips connected to \SI{500}{\micro\meter}~thick silicon sensors. Prior to analysis, tracks were pre-processed to remove halo effect~\cite{Manek2018_CTU}. If~$\left\{E_i\right\}_{i=1}^{n}$ labels deposited energies in each hit of the track, hits with energy below $0.626\max_{i=1}^n E_i + 103.75$ were removed.

In order to objectively evaluate proposed regression methods with ground truth information, artificial datasets were produced using the Allpix$^2$ framework~\cite{spannagel2018allpix2}. In simulated geometry, a Timepix3 detector was placed in an omnidirectional radiation field. Detector response was simulated for a variety of penetrating charged particles (listed in table~\ref{tab:sim-particles}). To eliminate angular bias due to the shape of the detector, data underwent probability-weighted sampling to ensure that the distribution of angles is indeed uniform. Furthermore, events with~$\theta<\ang{25}$ were removed to ensure that tracks have sufficient size.

\begin{table}[htbp]
\centering
\caption{\label{tab:sim-particles} Distribution of simulated primary events.}
\smallskip
\begin{tabularx}{0.55\textwidth}{|l|XX|}
\hline
Particle&Energy&Simulated events\\
\hline
Pion & \SI{40}{\GeV} & 81,013\\
Electron & \SI{600}{\MeV} & 49,349\\
Electron & \SI{10}{\MeV} & 11,292\\
Proton & \SI{300}{\MeV} & 1,689\\
Proton & \SI{200}{\MeV} & 1,647\\
\hline
\end{tabularx}
\end{table}

\begin{figure}
    %trim: left bottom right top
    \centering
    \includegraphics[height=4cm,trim={5 3 35 30},clip]{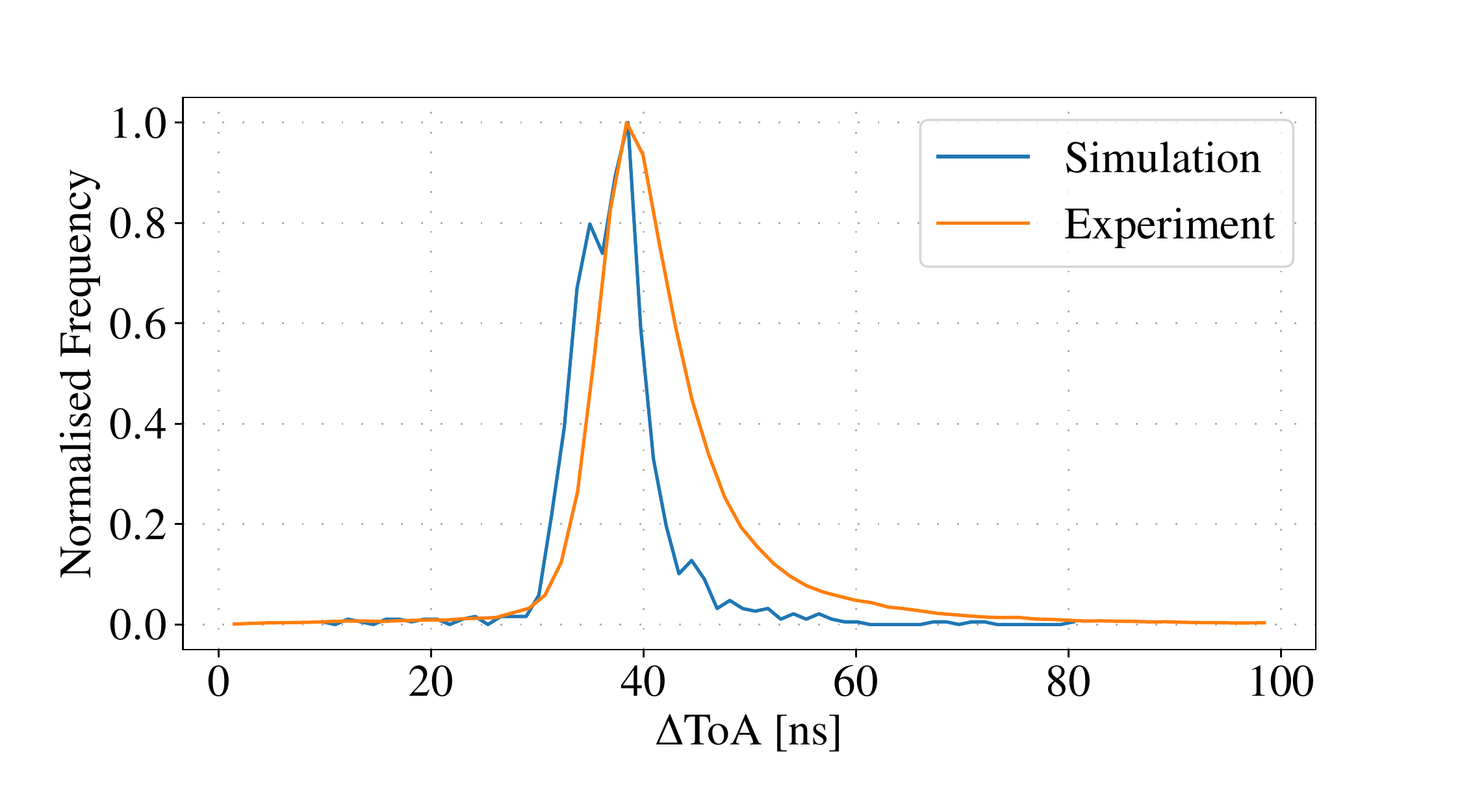}
    \includegraphics[height=4cm,trim={74 3 35 30},clip]{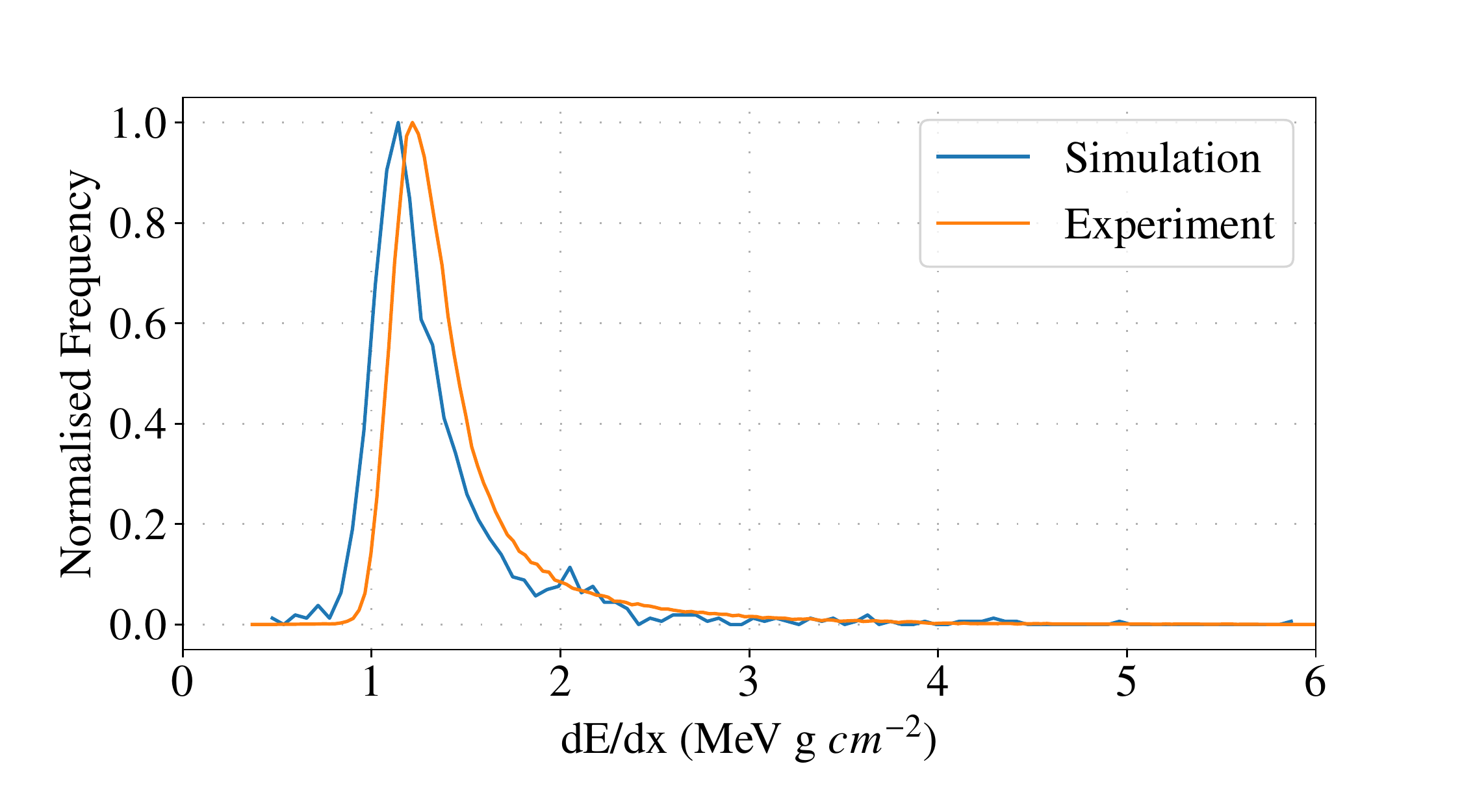}
    \caption{Comparison of simulation and real-world measurement. Plots show $\Delta$Time-of-Arrival (hit timestamp since the smallest timestamp within the cluster, left) and stopping power (right) calculated from datasets of pions tracks. In both plots, orange series represent 97,739~real events, whereas blue series correspond to 1,247~simulated events at equivalent energies. In both datasets the incidence angle~$\theta$ was~\ang{45} and the bias voltage was set to~\SI{150}{\volt}.}
    \label{fig:comparison-dedx-pions}
\end{figure}

To verify that simulated response is indeed realistic, artificial datasets were compared with real-world measurements (as shown in figure~\ref{fig:comparison-dedx-pions}), indicating overall agreement. Measured data were available from test beam campaigns at the Super Proton Synchrotron at CERN (pions of~\SI{40}{\GeV}) and from the Danish Center for Particle Therapy in Aarhus, Denmark\footnote{\url{https://www.en.auh.dk/departments/the-danish-centre-for-particle-therapy/}} (protons of various energies). Further details about the  measurements are provided later in dedicated sections.

\section{Comparison of methods}
\label{sec:comp}

\begin{figure}[tbp]
    \centering
    %trim: left bottom right top
    \includegraphics[height=4.5cm,trim={10 10 10 15},clip]{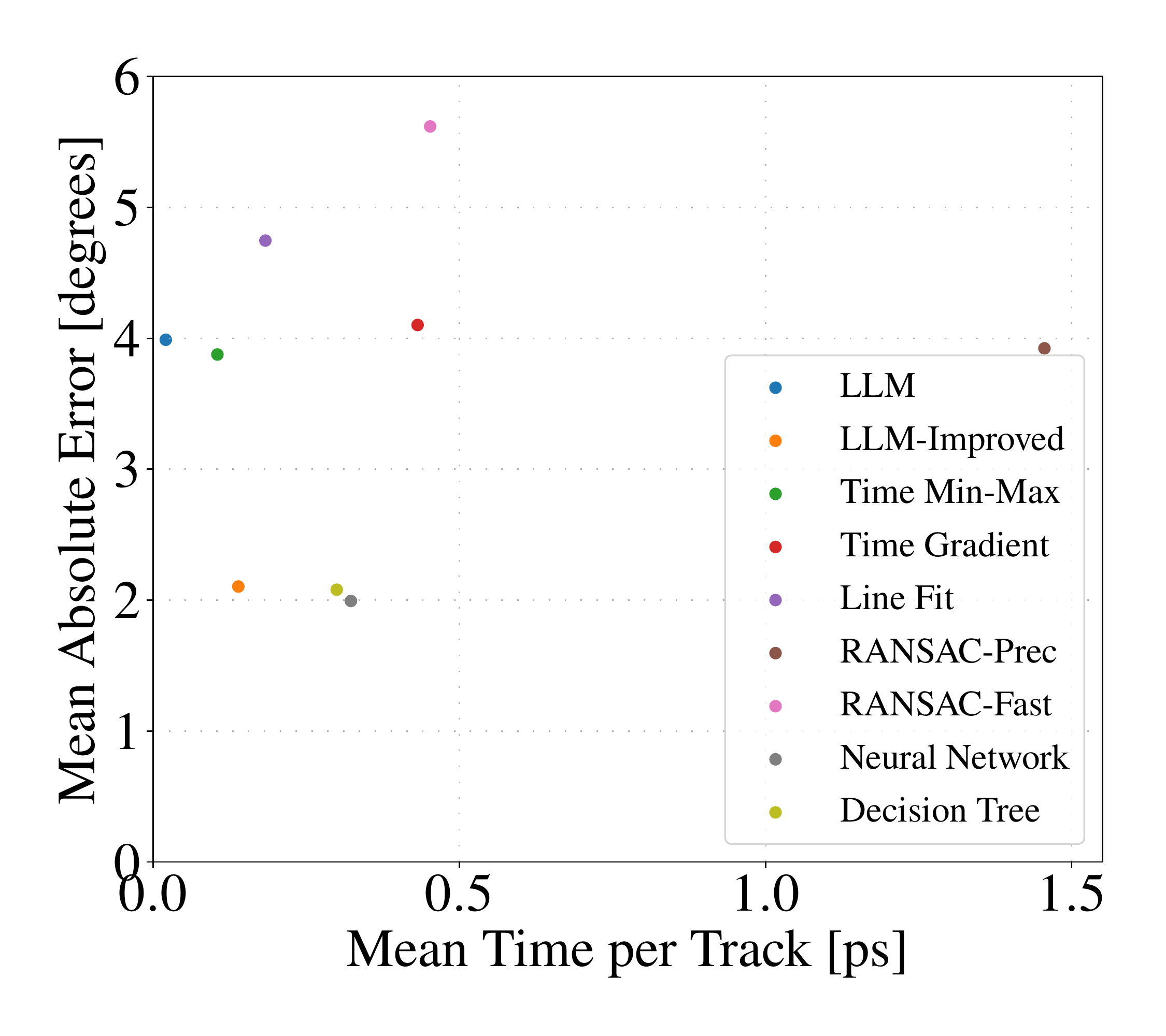}\hfill
    \includegraphics[height=4.5cm,trim={10 10 40 10},clip]{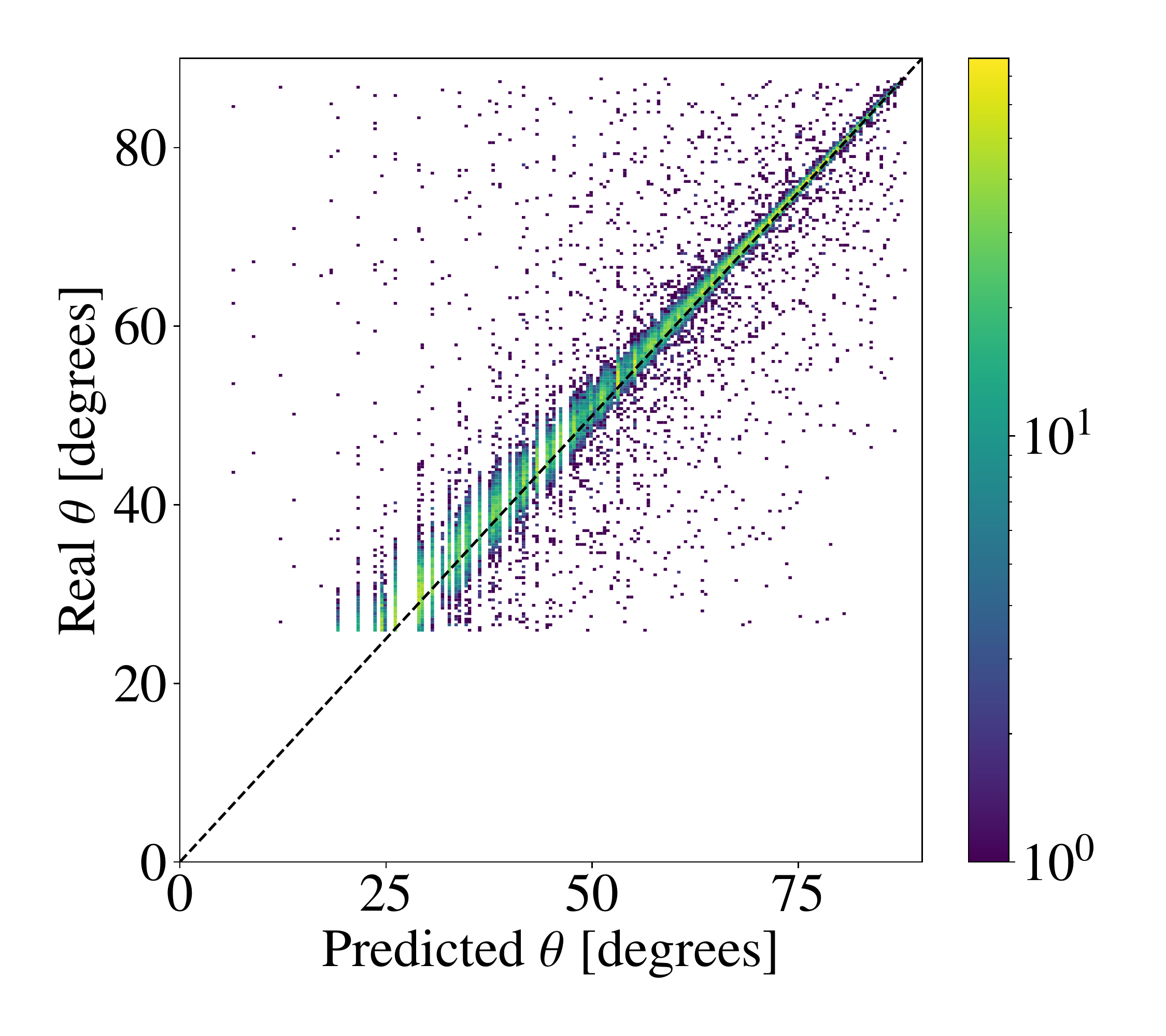}\hfill
    \includegraphics[height=4.5cm,trim={10 10 40 10},clip]{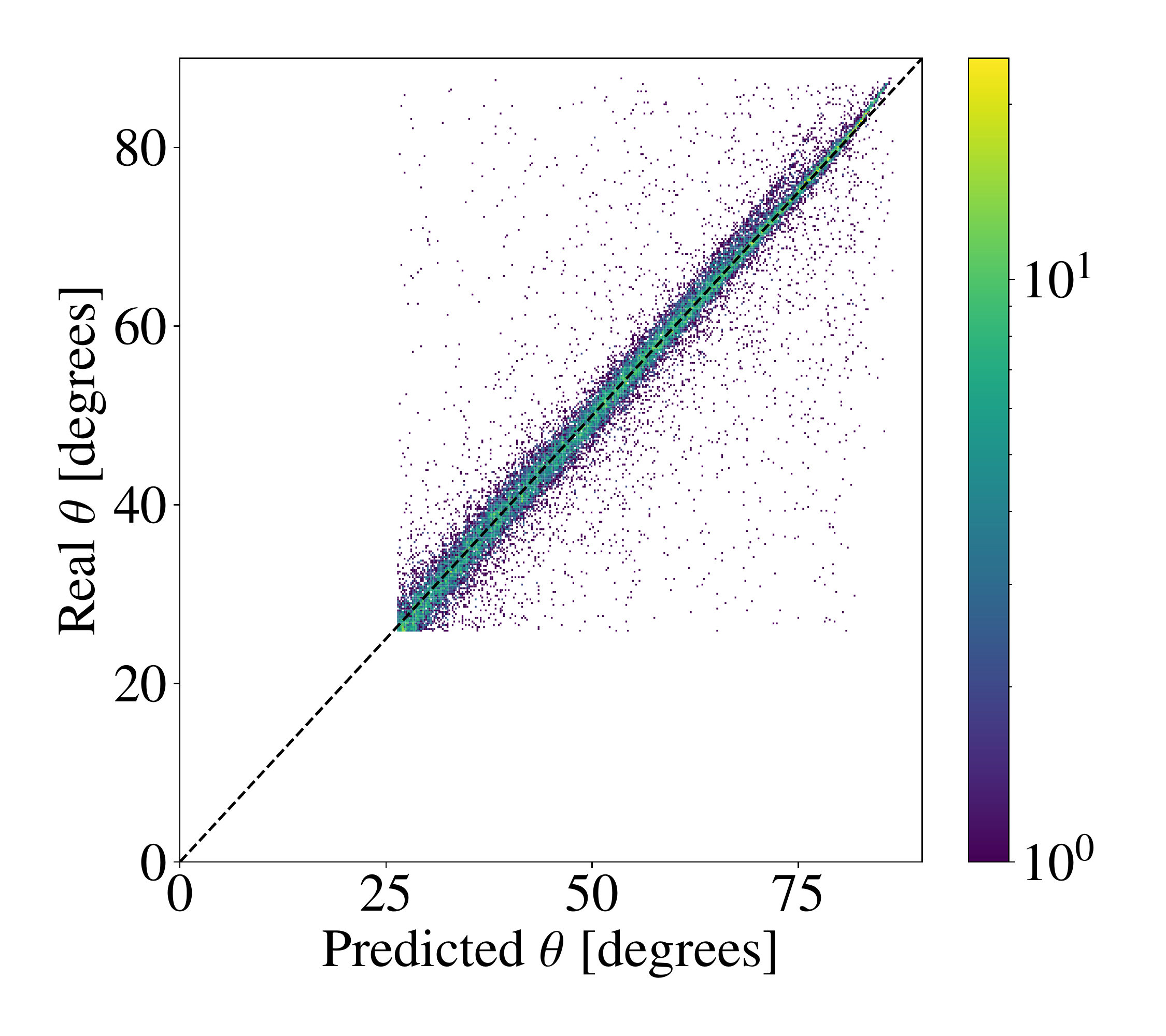}
    \caption{Comparison of regression methods for incidence angle~$\theta$. While the leftmost plot shows all methods' $E$ vs.~$\overline{t}$, the remaining plots show predicted vs.~true values produced by the most accurate methods in~$E$. From the left, LLM-Improved and neural network.}
    \label{fig:pred-true-theta}
\end{figure}

\begin{figure}[htbp]
    \centering
    %trim: left bottom right top
    \includegraphics[height=4.4cm,trim={10 0 10 60},clip]{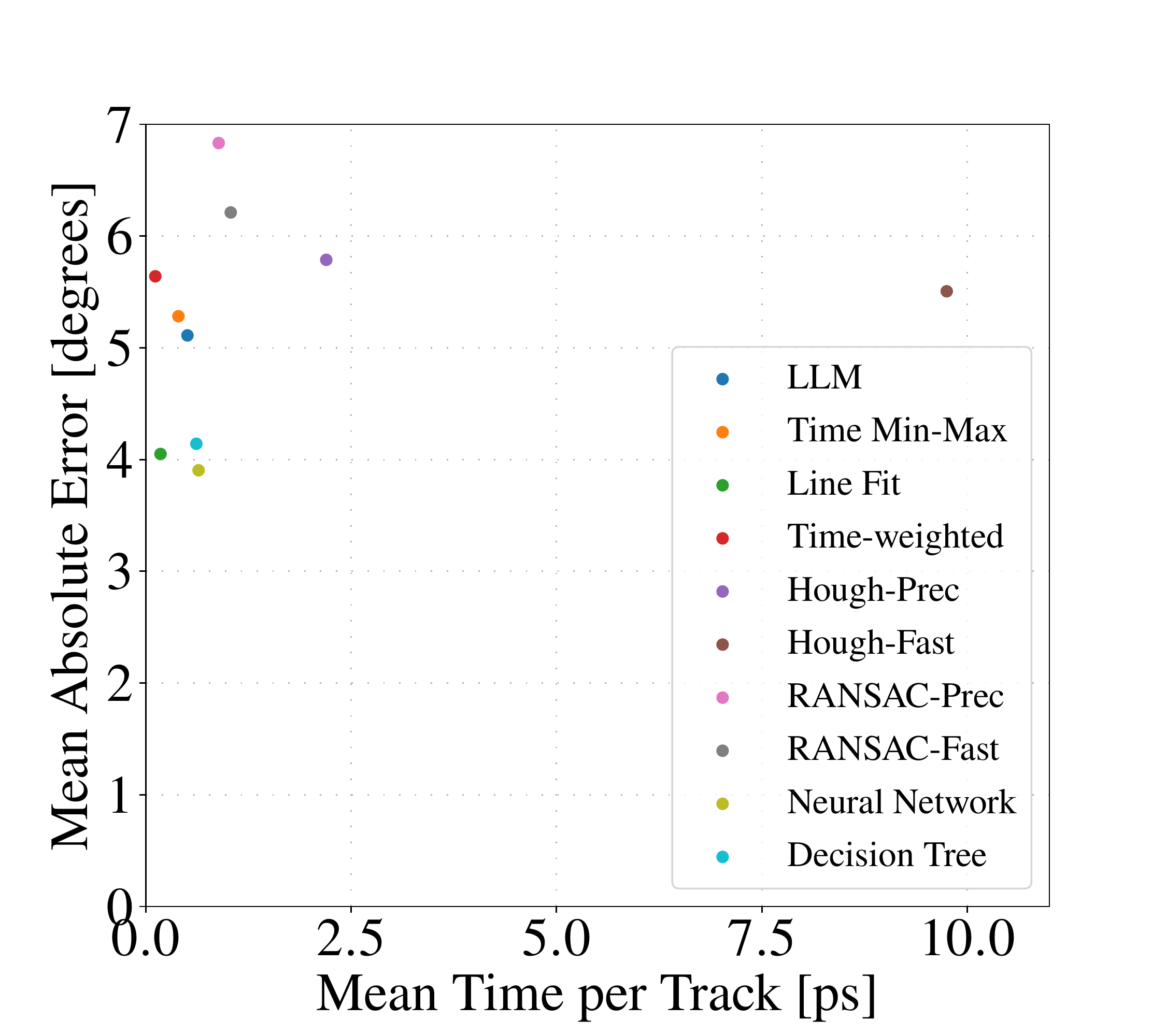}\hfill
    \includegraphics[height=4.4cm,trim={0 0 80 60},clip]{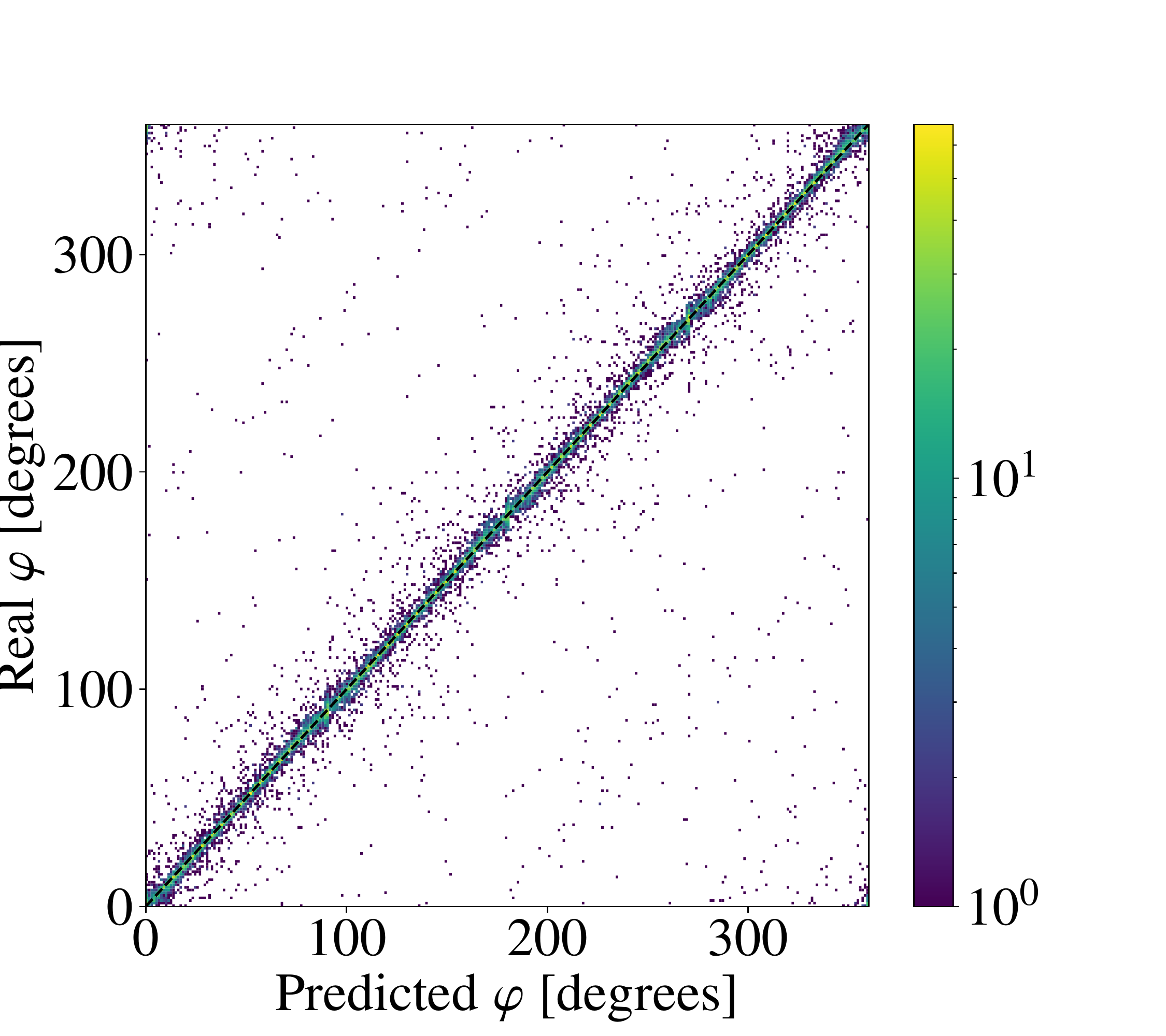}\hfill
    \includegraphics[height=4.4cm,trim={0 0 80 60},clip]{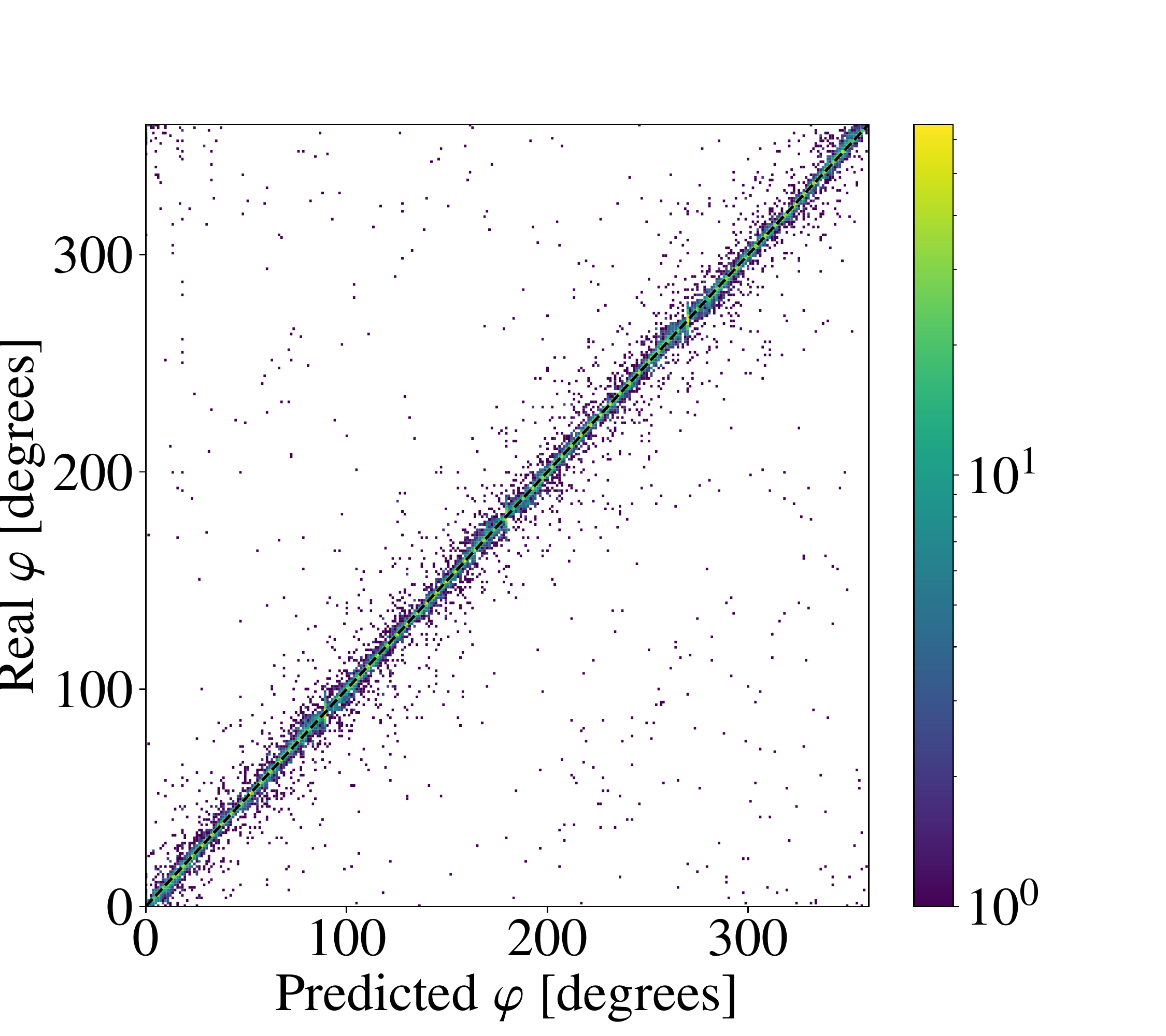}
    \caption{Comparison of regression methods for azimuth~$\varphi$, organized similarly to figure~\ref{fig:pred-true-theta}. Prediction plots show, from the left, line fit and neural network.}
    \label{fig:pred-true-phi}
\end{figure}

Presented methods were evaluated on simulated data according to the methodology outlined in the previous section. For each regression parameter, metrics were calculated independently. In addition, machine learning methods that require previous training (Neural Network, Decision Tree) used 80:20~split for training and testing set, respectively. Results of the evaluation are summarized in table~\ref{tab:comparison} and figures~\ref{fig:pred-true-theta}-\ref{fig:pred-true-phi}. We note that due to incompatibility or intractability, some methods were evaluated for a single regression parameter. Such methods are marked in table~\ref{tab:comparison} with dashes.

\begin{table}[htbp]
\centering
\caption{\label{tab:comparison} Results achieved by regression methods on simulated dataset, given in previously defined metrics. In each column, the best method's result is emphasized in bold.}
\smallskip
\begin{tabularx}{\textwidth}{|l|XXXX|XXXX|}
\hline
{}&\multicolumn{4}{c|}{Azimuth $\varphi$}&\multicolumn{4}{c|}{Incidence angle $\theta$}\\
Regression method&$\Delta$ [°]&$O$ [\%]&$R^2$&$\overline{t}$ [ps]&$\Delta$ [°]&$O$ [\%]&$R^2$&$\overline{t}$ [ps]\\
\hline
LLM & 8.19 & 2.93 & 0.74 & 0.29 & 3.99 & 3.85 & 0.72  & \textbf{0.02}\\
LLM-Improved & 5.11 & \textbf{1.48} & 0.80 & 0.38 & 2.10 & \textbf{2.41} & 0.91  & 0.14\\
Time Min-Max & 5.28& 1.60 & 0.81 & 0.40 &  3.88 & 3.13 & 0.78  & 0.11\\
Time Gradient & -- & -- & -- & -- & 4.10 & 3.01 & 0.75  & 0.43\\
Line Fit & 4.05 & 1.72 & 0.86 & 0.18  & 4.75 & 2.86 & 0.83  & 0.18\\
Time-weighted & 5.64 & 2.42 & 0.76 & \textbf{0.12} & -- & -- & --  & --\\
Hough-Fast & 5.78 & 1.78 & 0.77 & 2.20 & -- & -- & --  & --\\
Hough-Prec & 5.51 & 1.74 & 0.74 & 9.75 & -- & -- & --  & --\\
RANSAC-Fast & 6.83 & 1.49 & 0.66 & 0.89 & 5.62 & 3.60 & 0.74 & 0.45\\
RANSAC-Prec & 6.21 & \textbf{1.48} & 0.65 & 1.05 &  3.92 & 2.87 & 0.84 & 1.46\\
Neural Network & \textbf{3.90} & 1.61 & 0.80 & 0.64 & \textbf{1.99} & 2.61 & 0.91 & 0.32\\
Decision Tree & 4.14 & 1.60 & \textbf{0.86} & 0.62 & 2.08 & 2.79 & \textbf{0.92} & 0.30 \\
\hline
\end{tabularx}
\end{table}

For incidence angle~$\theta$, the most accurate estimations were produced by Neural Network with mean absolute error of~\ang{1.99}. The lowest outlier fraction~\SI{2.41}{\percent} was yielded by LLM-Improved and the largest~$R^2=\num{0.92}$ was achieved by Decision Tree. While according to all three metrics, these methods produced equivalent quality of predictions, their computational time differs. Specifically, machine learning models (Neural Network, Decision Tree) require approximately double time per single track to evaluate than LLM-Improved.

On the other hand, in spite of the achieved speed a certain degree of binning appears in the distribution of produced predictions (as seen in the middle plot of figure~\ref{fig:pred-true-theta}), especially for values of~$\theta$ below~\ang{45}. It is suspected that since this effect is only observed in two-stage regression methods that select hit pairs from tracks, it may be caused by detector pixelation. In small tracks that correspond to low incidence angles, the number of entry/exit point combinations limits possible predictions. In contrast, this effect is not observed in methods that are based on fitting or weighted means.

It is important to note at this point that the smallest mean time per track regardless of quality was achieved by the conventional LLM method with~$\overline{t}=\SI{0.02}{\pico\second}$. This represents a speedup of~$7\times$ in exchange for an error increase of~$2\times$ compared to the fastest and the most accurate of the previously discussed methods, respectively.

For azimuth~$\varphi$, the minimum error~\ang{3.90} was generated by Neural Network. The smallest percentage of outliers~\SI{1.48}{\percent} was produced by LLM-Improved as well as RANSAC-Prec, and the best $R^2=0.86$ was achieved by Decision Tree. Unlike the previous comparison, these methods do not seem to share similar behavior. We also note that overall accuracy for regression of the~$\varphi$ parameter seems to be generally lower compared to~$\theta$.

The fastest method regardless of quality was Time-weighted with mean time per track~$\overline{t}=\SI{0.12}{\pico\second}$. This constitutes a speedup of~$3\times$ in exchange for an error increase of~$1.5\times$ compared to the fastest and the most accurate of the previously discussed methods, respectively.

\section{Validation with real-world measurement}
\label{sec:real-world}

\begin{figure}
    \centering
    \includegraphics[width=0.45\linewidth]{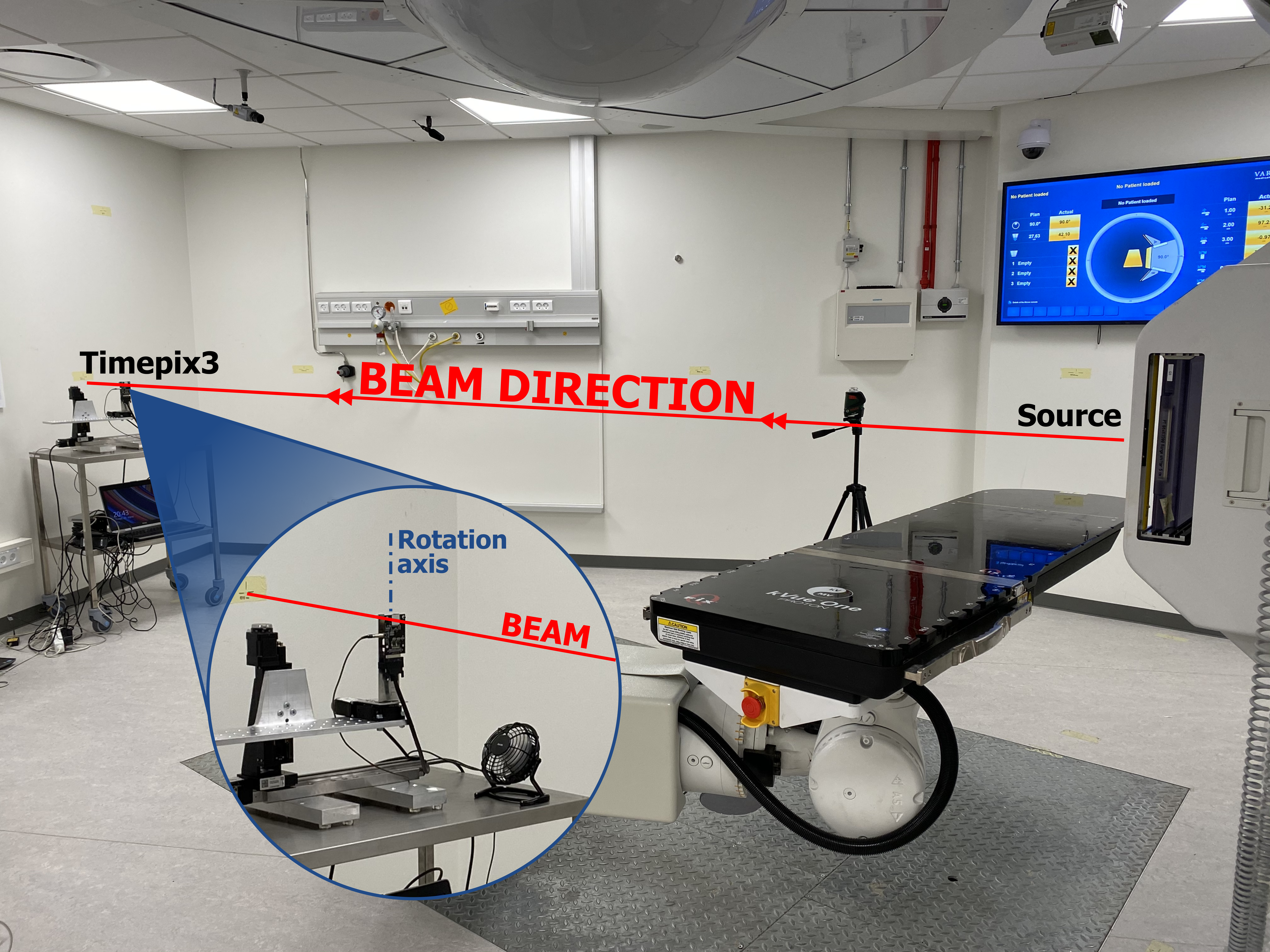}
    \caption{Hardware setup used to observe protons at the Danish Centre for Particle Therapy. Proton beam (red) is emitted from the right towards a Timepix3 detector in the left. A detail (lower left) indicates detector's relative to beam.}
    \label{fig:denmark-placement}
\end{figure}

Following their comparison, regression methods were applied to experimental data measured in a radiation field of known characteristics. The dataset comprises proton tracks measured in August~2020 at the Danish Centre for Particle Therapy. A Timepix3 detector was placed on a rotating motorized stage and positioned directly in the path of proton beam (as shown in figure~\ref{fig:denmark-placement}). The chip was irradiated by protons of energies 125, 171 and~\SI{219}{\MeV}, and rotated along the indicated axis to face the beam at angles \ang{30}, \ang{60} and \ang{75}.

Based on results of the comparison in previous section, two regression methods were selected for this analysis. The incidence angle~$\theta$ was predicted by LLM-Improved, which achieved the smallest outlier percentage. In addition, the method was simple to implement and proved to be computationally tractable. The azimuth~$\varphi$ was estimated using Line Fit. Even though this method was superior relative to others in any attribute, it was selected because it represents a good compromise between tractability and accuracy.

\begin{figure}
    \centering
    %trim: left bottom right top
    \includegraphics[height=5.5cm,trim={25 20 30 40},clip]{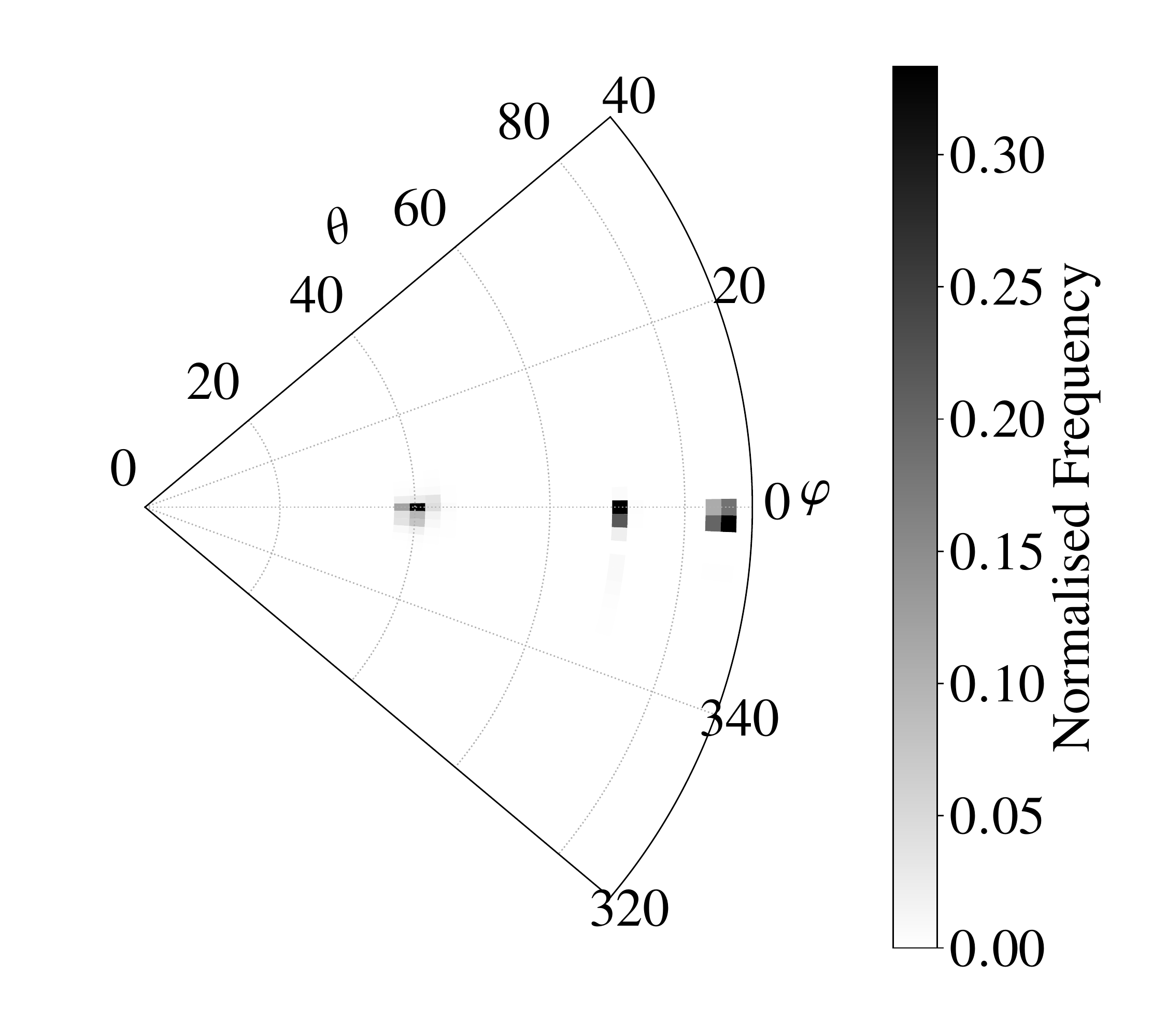}\hspace{0.5cm}
    \includegraphics[height=5.5cm,trim={20 0 60 65},clip]{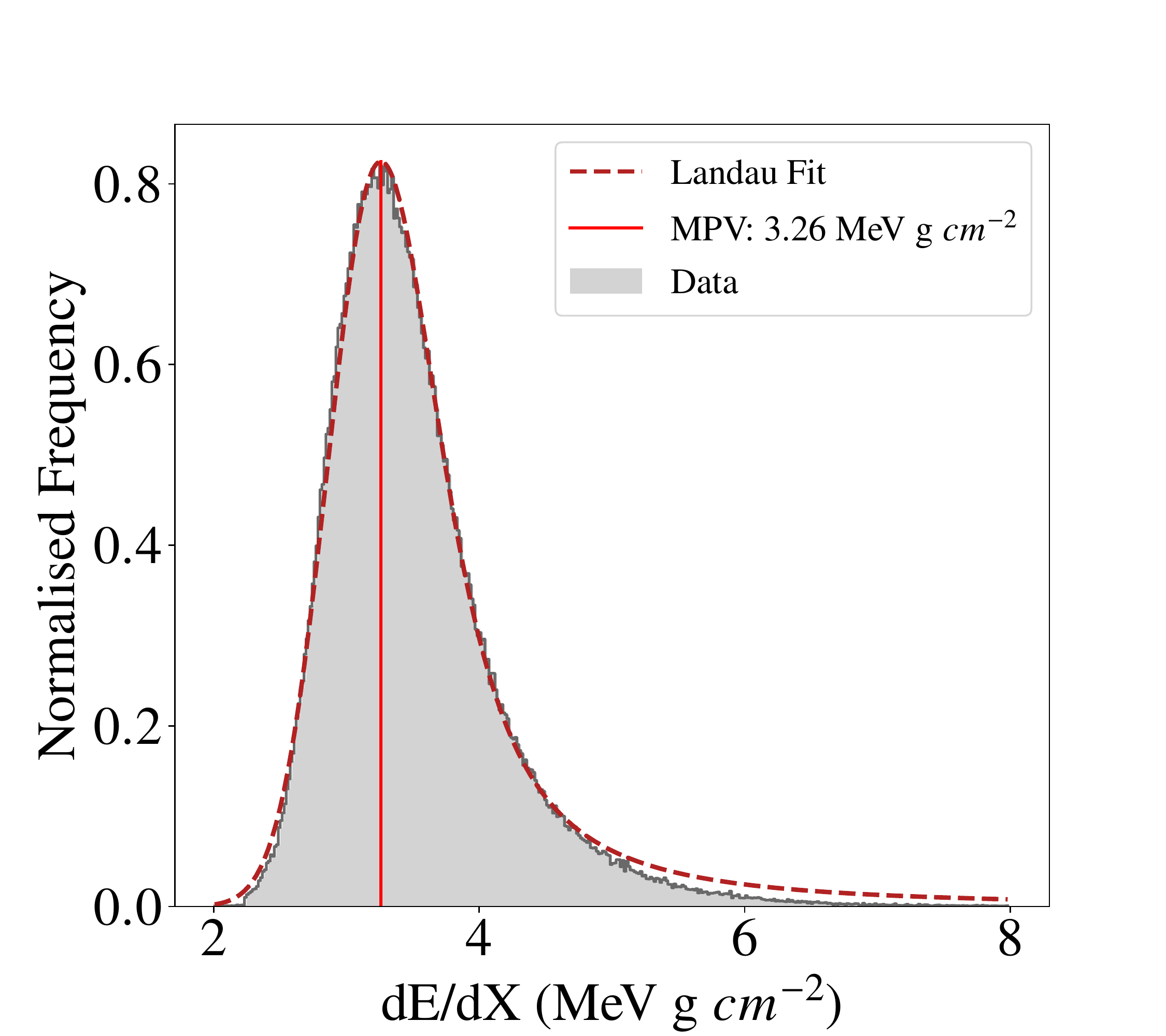}
    \caption{Results of selected methods operating on \SI{218.51\pm 0.95}{\MeV}~proton tracks measured at the Danish Centre for Particle Therapy. While the polar plot (left) is a 2D histogram of predicted regression parameters, the histogram (right) shows stopping power calculated based on those predictions. A fitted Landau distribution is superimposed.}
    \label{fig:denmark-polar}
\end{figure}

Results are summarized in figure~\ref{fig:denmark-polar}. In the angular map (left), a series of peaks emerges at~$\varphi\approx\ang{0}$ and various values of~$\theta$. This is consistent with expectation, given that rotation of the detector changes~$\theta$, but has no direct effect on~$\varphi$. In addition, locations of peaks approximately correspond to programmed impact angles. Finally, regression outputs were used to calculate stopping power $\mathrm{d}E/\mathrm{d}X$ (given in the right plot of figure~\ref{fig:denmark-polar}) that was fitted with a Landau distribution with most probable value of~\SI{3.26\pm 0.11}{\MeV\g\per\cm\squared}.

\section{Application to radiation fields with unknown characteristics}
\label{sec:exp}

In this section, we aim to offer broader comparison of regression methods with other alternatives. To this end, our methodology is applied to datasets measured at the Large Hadron Collider that were analyzed in previous studies. Physics interpretation is out of scope of this work.

\begin{figure}
    %trim: left bottom right top
    \centering
    \includegraphics[height=5.5cm,trim={25 20 30 25},clip]{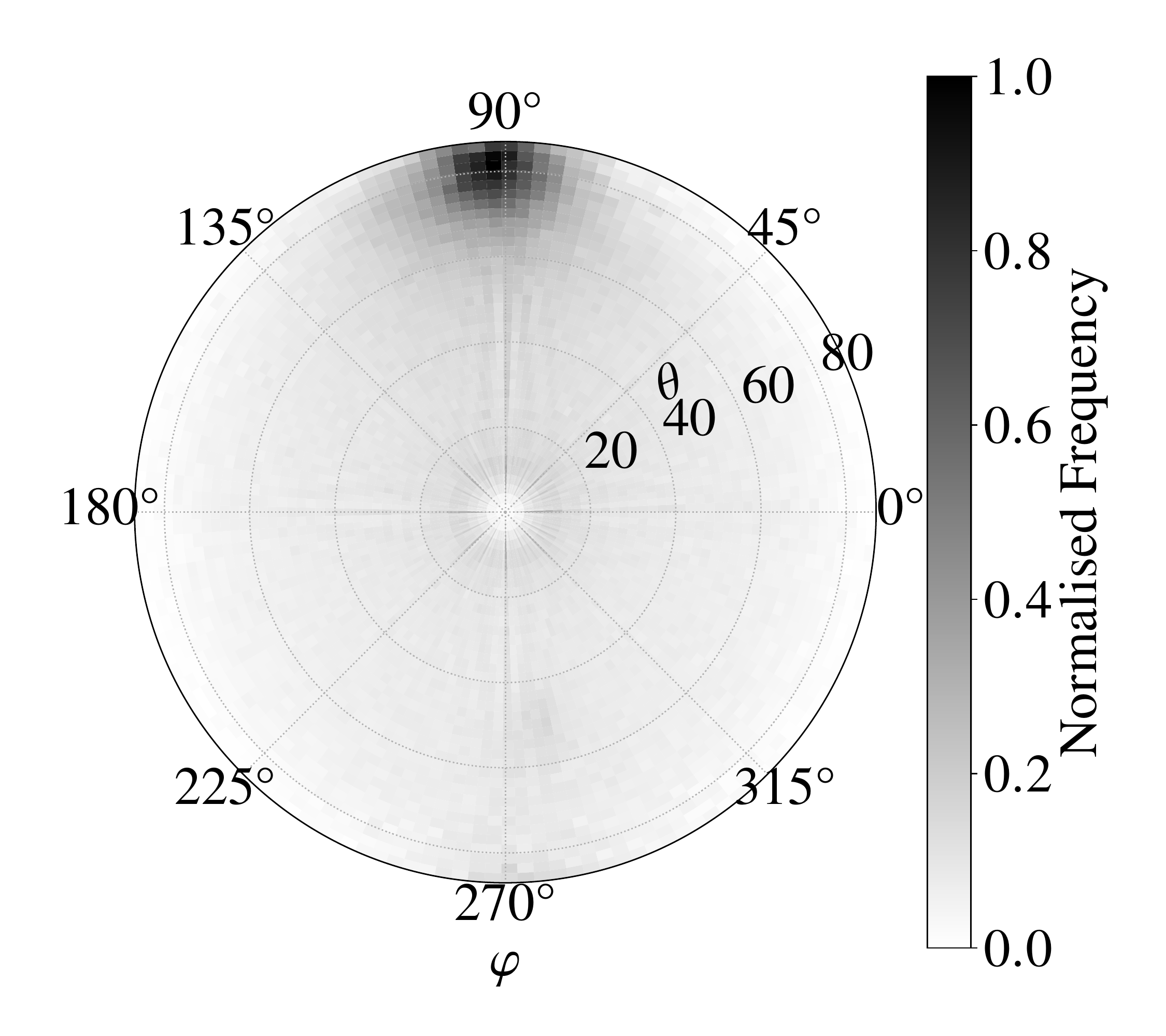}\hspace{0.75cm}
    \includegraphics[height=5.5cm,trim={0 0 60 65},clip]{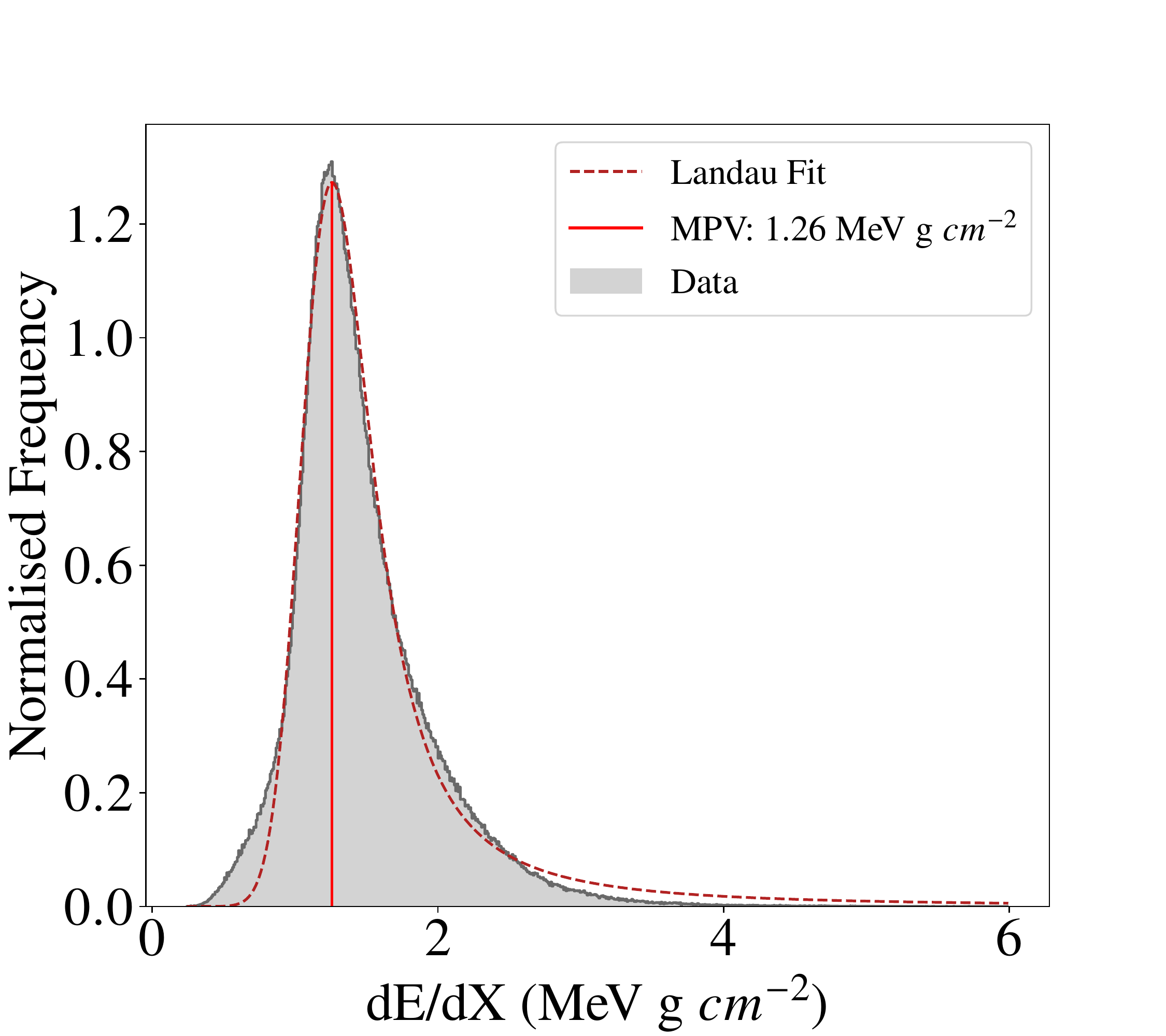}\\
    \includegraphics[height=5.5cm,trim={25 20 30 25},clip]{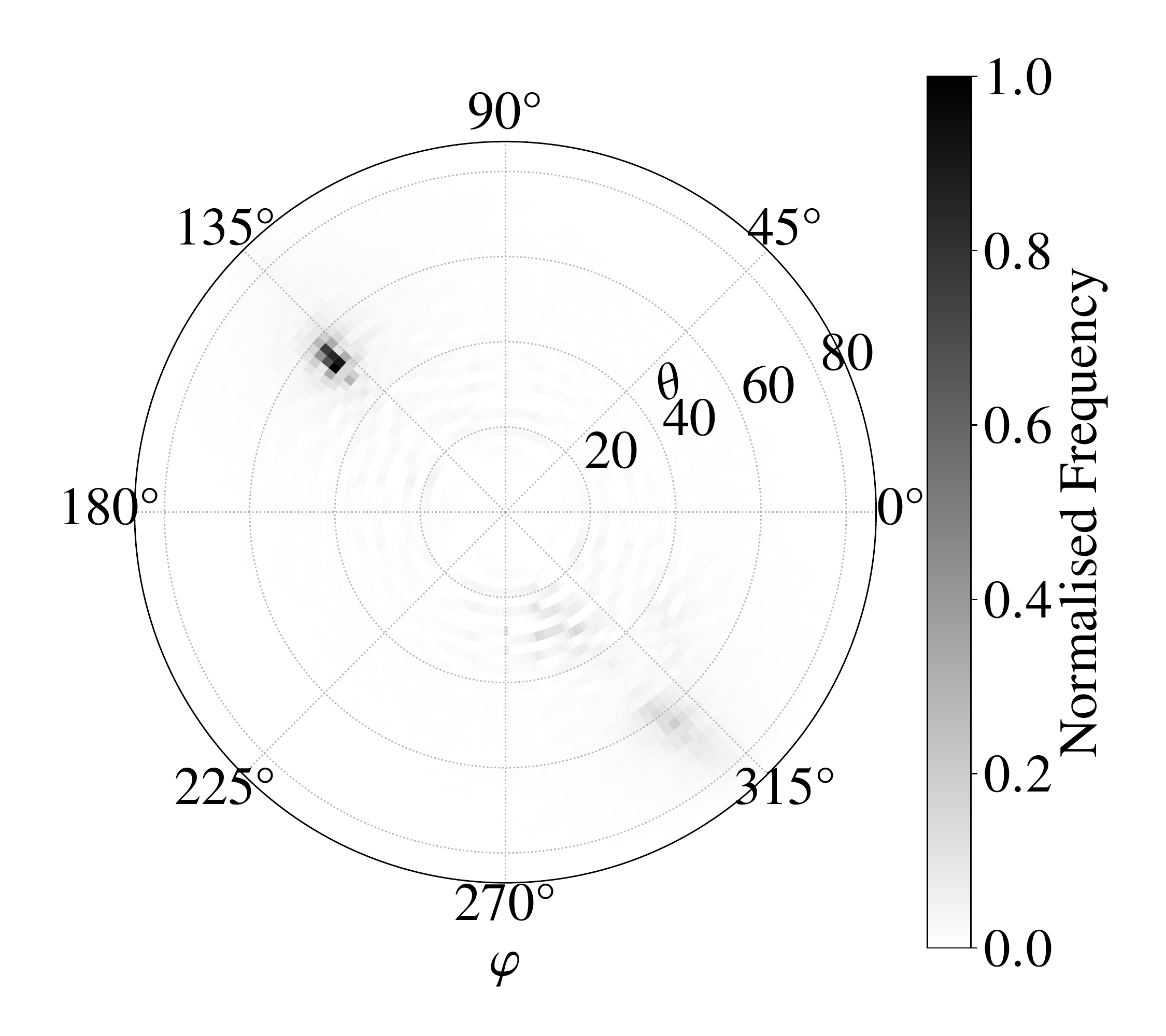}\hspace{0.75cm}
    \includegraphics[height=5.5cm,trim={0 0 60 65},clip]{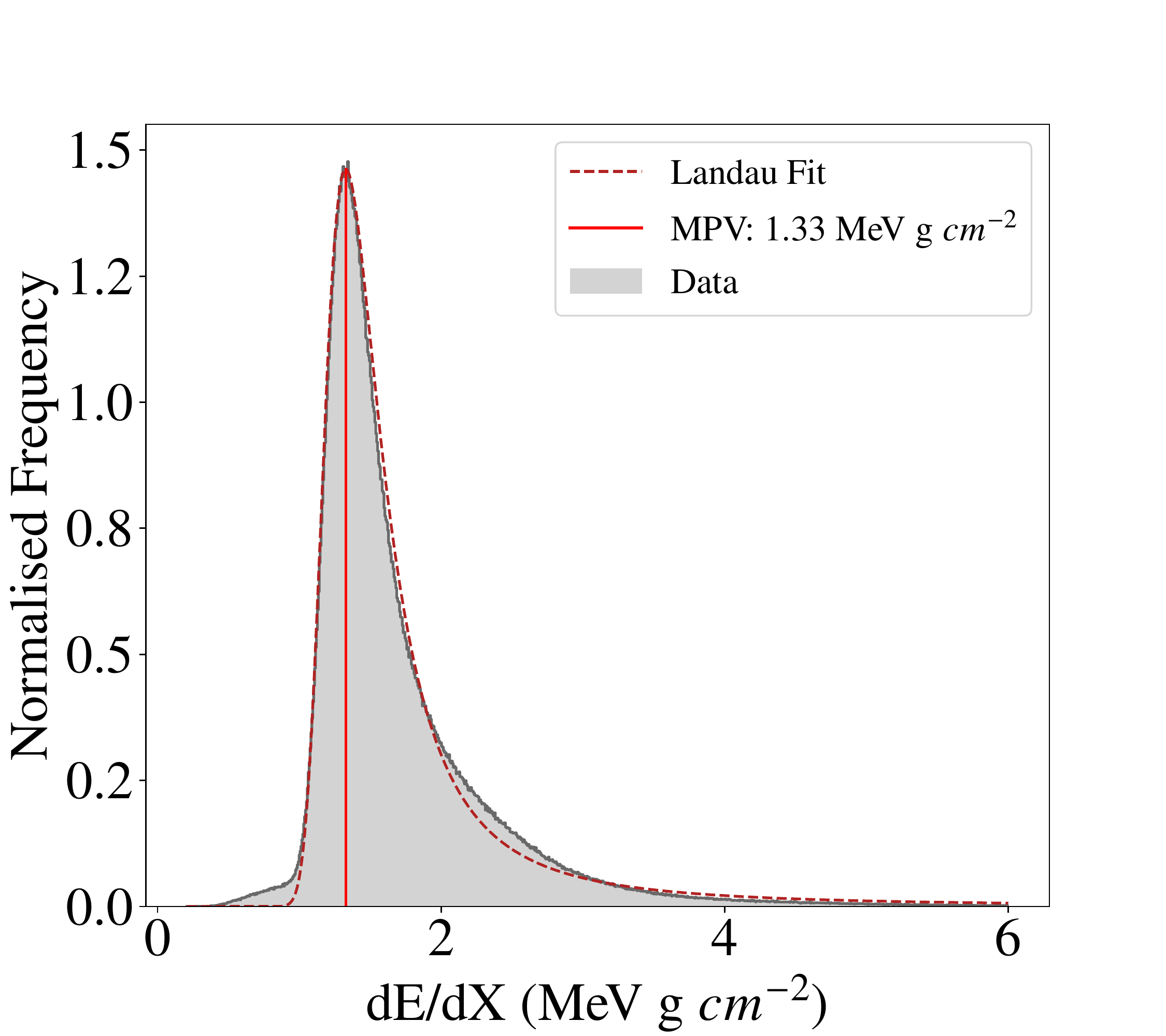}
    \caption{Results from ATLAS (top) and the MoEDAL experiment (bottom), organized similar to figure~\ref{fig:denmark-polar}.}
    \label{fig:atlas-moedal-polar}
\end{figure}

The first presented dataset was generated by a Timepix3 detector installed in ATLAS~\cite{bergmann2020particle} at $(X,Y,Z)=(\SI{-3580}{\mm},\SI{970}{\mm},\SI{2830}{\mm})$, taking data on September 7, 2018. Due to relatively high frequency of small incidence angles, regression methods that were shown to be stable under such conditions were selected. Azimuths~$\varphi$ were predicted by Time Gradient, whereas incidence angles~$\theta$ were estimated by the Time-weighted method. Furthermore, tracks of total energy below~\SI{180}{\keV} were excluded. Results are displayed in the top row of figure~\ref{fig:atlas-moedal-polar}. The angular map (left) shows distributed response with a distinct peak at~$(\varphi,\theta)=(\ang{90.01},\ang{75.9})$. The stopping power spectrum (right) was fitted with a Landau distribution with most probable value of~\SI{1.32\pm 0.73}{\MeV\g\per\cm\squared}.

The second dataset comes from the MoEDAL experiment~\cite{acharya2021timepix3}, where two Timepix3 detectors were installed at a distance of~\SI{1.1}{\meter} from the LHCb interaction point. Data taken on November 12, 2018 were analyzed by the methods used in section~\ref{sec:real-world}, producing results that are shown in the bottom row of figure~\ref{fig:atlas-moedal-polar}. In the angular map (left), a narrow peak at~$(\varphi,\theta)=(\ang{138.97},\ang{54.06})$ can be identified. While there is only little material in between the interaction point and Timepix3 at MoEDAL, in ATLAS the device is farther away from the interaction point and particles have to penetrate several layers of additional material. Thus, trajectories of particles generated at the interaction point are affected by small angle scattering and secondary particles create omnidirectional background. The stopping power spectrum (right) was fitted with a Landau distribution with most probable value of~\SI{1.25\pm 0.10}{\MeV\g\per\cm\squared} that most likely corresponds to minimum ionizing particles (MIPs).

\section{Conclusion}
\label{sec:conclusion}

In this work, 10~methods for estimating particle directions from Timepix3 tracks were proposed. Firstly, these methods were evaluated on simulated datasets and rated by their accuracy and computational tractability. Secondly, two selected methods were applied to a real-world dataset that was measured in radiation environment of known characteristics. Finally, selected methods were applied to experimental data that describe previously studied environments, which are not yet fully understood.
In all instances, presented methods performed consistently with expectations. Under simulated conditions, the most accurate method for both regression parameters was Neural Network, which achieved mean absolute error of~\ang{3.90} and~\ang{1.99} for~$\varphi$ and~$\theta$ respectively. In a real-world measurement where predominant particle directions were known, peaks were identified near expected locations and in expected quantities. In experimental datasets, produced results agreed with those of previous studies.

In the future, presented regression methods would benefit from a more extensive evaluation on larger datasets. In addition to upscaling, more valuable information may be obtained by testing the methods with other particle species and energies. Here of particular interest are electrons at lower energies than~\SI{10}{\MeV} and heavy ionizing particles (for instance, ions or ion-like tracks).
\section*{Acknowledgements}
\label{sec:acknowledgements}

This work was carried out within the Medipix collaboration.

% this makes all un-typeset figures appear
\FloatBarrier

\end{document}